\documentclass[12pt]{article}
\usepackage{color,amsmath,amsfonts,amssymb,a4,epsfig}

\newcommand{\R}{{\mathbb{R}}}
\newcommand{\C}{{\mathbb{C}}}

\newcommand{\E}{{\mathbb{E}}}
\newcommand{\D}{{\mathbb{D}}}

\def\ha{\frac{1}{2}}

\def\pa{\partial}
\def\ra{\rightarrow}
\def\preuve{\begin{proof}} 
\def\ga{\alpha}
\def\gb{\beta}

\def\gd{\delta}
\def\ge{\varepsilon}

\def\gg{\gamma}

\def\gl{\lambda}
\def\go{\omega}

\def\gs{\sigma}

\def\OPD{${\rm \Psi}$DO}

\def\ppal{principal~}
\newtheorem{defi}{Definition}

\newtheorem{lemm}{Lemma}
\newtheorem{prop}{Proposition}
\newtheorem{rem}{Remark}
\newtheorem{coro}{Corollary}
\newtheorem{theo}{Theorem}
\newtheorem{exem}{Example}[section]
\newenvironment{demo}{\noindent {\it Proof.--}
      \begin{quotation}\noindent}{\end{quotation}\hfill$\square $}

\begin{document}

\title{Mathematical models for passive imaging \\
I: general background.}
\author{Yves Colin de Verdi\`ere \footnote{Institut Fourier,
 Unit{\'e} mixte
 de recherche CNRS-UJF 5582,
 BP 74, 38402-Saint Martin d'H\`eres Cedex (France);
\textcolor{blue}{
 http://www-fourier.ujf-grenoble.fr/${\rm \tilde{~} }$ycolver/  }~
}}


\maketitle

\begin{abstract}
Passive imaging is a new technics which has been proved to be very efficient,
for example in seismology: the correlation of the noisy fields
between different points is strongly related to the Green function
of the wave propagation. The aim of this paper is to provide
a mathematical context for this approach and to show, in particular,
how the methods of semi-classical analysis can be be used in order to
find the asymptotic behaviour of the correlations.

\end{abstract}

\section*{Introduction}
Passive imaging is a way to solve inverse problems: it has been succesfull
in seismology and  acoustics \cite{DL,DLC,LW,SC,SCSR,WL1,WL2,We}.
The method is as follows: let us 
assume that we have a medium $X$ (a smooth manifold) 
and a smooth,
deterministic (no randomness in it) linear wave equation in $X$.
We hope  to recover  (part of) the  geometry of $X$ from the wave propagation.
We assume  that there is somewhere in $X$ a source of noise ${\bf f}(x,t)$
which is a stationary random field. 
This source generates, by the wave propagation, a field ${\bf u}(x,t)=
\left( u^\ga (x,t) \right)_{\ga =1,\cdots, N}$
which people do record on long time intervalls.
We want to get some information on the 
propagation of waves from $B$ to $A$ in $X$ from the correlation
matrix\footnote{For every matrix
$(a_{ij})$, we write $(a_{ij})^\star:=(\overline{a_{ji}})$.} 
\[ C_{A,B}(\tau)=\lim  _{T\ra \infty}\frac{1}{T}\int _0^T {\bf u}(A,t)
\otimes {{\bf u}(B,t-\tau)}^\star dt~
\]
(equivalently 
\[ C_{A,B}^{\ga\gb}(\tau)=\lim  _{T\ra \infty}\frac{1}{T}\int _0^T {\bf u}^\ga(A,t)
 {\overline{{\bf u}^\gb (B,t-\tau)}} dt~)
\]
 which can be computed numerically from the  fields recorded 
 at $A$
and $B $.
It turns out that $C_{A,B}(\tau) $ is closely related to the
deterministic Green's function $G(A,B,\tau)$ of the wave equation in $X$.
It means that one can hope to recover, using Fourier analysis,
 the propagation speeds of waves
between $A$ and $B$ as a function of the frequency, or, in other words,
the so-called dispersion relation.

If the wave dynamics is time reversal symmetric,
the correlation  admits also a symmetry by
change of $\tau $ into  $-\tau$; this observation
has been used for  clock synchronization in the ocean, see \cite{SRTDHK}.

The goal of this paper  is to give precise formulae for
$C_{A,B}(\tau) $ in the high frequency limit assuming a 
rapide decay of correlations of the source ${\bf f}$.
More precisely, we  have 2 small parameters, one of them entering
into the correlation distance of the source noise, the other one 
in the high frequency propagation. The fact that both are of the same
order  of magnitude is crucial for the method.

Let us also mention on the technical side that, rather than  using  mode 
decompositions, we prefer to work  directly with the dynamics; 
in other words,  we need really a {\it time dependent} rather than
 a  {\it stationary
approach}. Mode decompositions  are  often usefull, but they are 
 of no much help
for general operators with no particular symmetry.

For clarity, we will first discuss the non-physical case
of a first order wave equation like the Schr\"odinger equation, then the
case of a more usual wave equations (acoustics, elasticity).

The main  result expresses, for $\tau >0$,
  $C_{A,B}(\tau) $ as the Schwartz
kernel of 
$ \Omega (\tau )\circ \Pi  $ where $\Pi $
 is a suitable pseudo-differential operator (a ~\OPD),
whose principal symbol can be explicitely computed,
  and $\Omega (\tau )$
is the  (semi-)group of the (damped) wave propagation.
It implies that we can recover the dispersion relation,
i.e. the classical dynamics,  from the knowledge of
all two-points correlations.

In order to make  the paper  readable  by a large set of people,
we have tried to make it self-contained by including
 sections on pseudo-differential operators and
on random fields.

In Section \ref{sec:general}, we start with a quite general setting
and discuss a general formula for the correlation (Equation
(\ref{equ:correl})).

Section \ref{sec:algeb} is devoted to exact formulae in case of an
homogeneous white noise.

In Section \ref{sec:TRS}, we discuss the important property
of time reversal symmetry which plays a prominent part  in the  applications
and is also usefull   as a numerical test.

In Section \ref{sec:random}, we introduce a large family of anisotropic random
fields and show the relation between their power spectra and
the Wigner measures.

Section \ref{sec:pseudos} contains the main result expressing the correlation
in the case of a Schr\"odinger wave equation.

Section \ref{sec:wave} does the same in case of a wave equation.

The short Section \ref{sec:problems} is a  problem section.

Section \ref{sec:scattering} is a about a quite independent issue
relative to correlations of scattered waves.

Finally, there is a very short introduction to pseudo-differential operators
in Section \ref{sec:psido}.

{\it Acknowledgment: I would like to thank {\rm Michel Campillo }
and his colleagues from LGIT for discussions and collaborations;
thank you also to {\rm Raoul Robert} for the starting ideas about 
constructing random fields.}

\section{A general formula for the correlation} \label{sec:general}

\subsection{The model}
We will first consider the following damped wave equation:
\begin{equation}\label{equ:PI}
  \frac{d{\bf u}}{dt}+ \hat{H} {\bf u}= {\bf f}
\end{equation}

\begin{itemize}
\item $X$ is a {\it smooth manifold} of dimension $d$ 
with a smooth measure $|dx|$
\item ${\bf u}(x,t),~x\in X,~t\in\R $ is  {\it the field }(scalar or vector
valued)
with values in $\C^N$ (or $\R^N$).
\item The linear operator 
$\hat{H}$ is the  {\it Hamiltonian,} acting  on
 $L^2(X,\C ^N)$. It  satisfies some attenuation property:
if we define the semi-group $\Omega (t)={\rm exp}(-t\hat{H}),t\geq 0$,
there exists $k>0$, so that 
we have the estimate  $\| \Omega (t) \| =0(e^{-kt})$.
\item The  {\it source} $ {\bf f}$ is a {\it stationary ergodic random field }
on $X\times \R $ with values in $\C^N$ (or $\R^N$) 
whose matrix valued correlation kernel
is given
 by
\begin{equation} \label{equ:correl_source}
 \E ( {\bf f}(x,s)\otimes  {\bf f}^\star (y,s'))=
K(x,y,s-s') ~.\end{equation}
We will usually assume that $K(x,y,t)$ vanishes for 
large $t$, say $|t|\geq t_0 >0$.
\end{itemize}

\subsection{Examples}
\subsubsection{Schr\"odinger equation}
Let $X$ be a smooth Riemannian manifold with Laplace-Beltrami  operator
$\Delta $.
Let us give $a:X\ra \R $ a non negative  function,
$V$ a smooth real valued function on $X$, $\hbar $ a non negative
constant,  and
$ \hat{K}=-\hbar^2 \Delta + V(x) $, and take:
\[ \frac{\hbar}{i}(u_t +a(x)u)  + \hat{K}u= g~. \]
It is a particular case of Equation (\ref{equ:PI})
where $\hat{H}= \frac{i}{\hbar} \hat{K}+ a(x)  $
and $f=\frac{i}{\hbar}g$. Let us note for future use that, if $\hbar \ra 0$, 
the principal symbol of our equation is
$\omega + \| \xi \|^2 + V(x)$
and $a(x)$ is only a sub-principal term entering into the
transport equation, but not in the classical dynamics.

\subsubsection{Wave equations}\label{subsubsection:wave}
Let us start with 
\begin{equation}\label{equ:wave}
 u_{tt}+2a u_t -\Delta u= f, ~a> 0  \end{equation}
(with $\Delta $ the Laplace-Beltrami  operator of a Riemannian metric on $X$)
which corresponds to Equation (\ref{equ:PI}) with 
\[ {\bf u}=
\left( \begin{array} {c} u \\ u_t \end{array} \right),~
{\bf f}:=\left( \begin{array} {c}0  \\ f \end{array} \right)~.
 \]
and
\[ \hat{H}=
\left( \begin{array} {cc} 0 &-{\rm Id} \\ -\Delta & 2a \end{array} \right)
 ~.\]

\subsubsection{Pseudo-differential equations}

We can assume that the dynamics is generated by a ~\OPD~
 (see Section \ref{sec:psido}).
Our equation looks then like:
\[ \frac{\ge}{i}{\bf u}_t +\hat{H}_\ge {\bf u}={\bf f} \]
with
\[ \hat{H}_\ge ={\rm Op}_\ge (H_0+\ge H_1) ~.\]

 This allows to include
\begin{itemize}
\item An effective surface Hamiltonian associated to stratified media
(included in the $H_0$ term)
 \cite{YCdV2}. They are usually ~\OPD's with a non trivial 
dispersion relation
\item Frequency dependent damping included in $H_1$:
 this is usually the case for
seismic waves.
\end{itemize}

\subsection{The correlation}
\begin{defi}\label{defi:prop}
Let us define, for $t\geq 0$,
 $\Omega(t):={\rm exp}(-t\hat{H})$ and the {\rm propagator} $P$
by the formula: 
\[ (\Omega(t){\bf v})(x)=\int _X P(t,x,y){\bf v}(y) |dy|~. \]
\end{defi}

 The propagator $P$ satisfies
\[ \int _X P(t,x,y)P(s,y,z)|dy| =P(t+s,x,z) \]
which comes from:
$\Omega(t+s)=\Omega(t)\circ \Omega(s)$.
The causal solution of Equation (\ref{equ:PI}) is then given by
\[ {\bf u}(x,t)=\int_0^{\infty}ds 
 \int_X P(s,x,y){\bf f}(t-s,y)|dy| ~.\]

We define 
\[ C_{A,B}(\tau):=\lim_{T\ra +\infty} \frac{1}{T}\int_0^T {\bf u}(A,t)\otimes
 {\bf u}(B,t-\tau)^\star dt ~.\]

Ergodicity allows to replace time average by ensemble average
and  we get by a simple calculation:
\begin{theo}
If $P$ is defined as in Definition \ref{defi:prop}
and $K$ by Equation (\ref{equ:correl_source}),
  we have, for $\tau >0$: 
\begin{equation} \label{equ:correl}
 C_{A,B}(\tau )=\int _{0}^{\infty }
ds\int_{-\infty}^{s+\tau}d\gs
 \int _{X\times X} |dx||dy| 
P(s+\tau-\gs  ,A,x)K(x,y,\gs )P^\star (s,B,y)
\end{equation}
and $C_{A,B}(-\tau)=C_{B,A}(\tau )^\star $.

We get, for $\tau > t_0$,  the  formula\footnote{If $R$ is an operator 
we will denote by $[R](x,y)$ its Schwartz kernel; $\hat{L}$ is the
 integral operator whose kernel is $L(x,y)$}:
\begin{equation} \label{equ:corr_oper}
 C_{A,B}(\tau )= [  \Omega(\tau)\Pi ](A,B) 
\end{equation}
with 
\begin{equation} \label{equ:Pi}
 \Pi= \int _0^\infty \Omega (s) {\cal L}\Omega ^\star (s) ds \end{equation}
and
\[  {\cal L}=\int _{|t|\leq t_0} \Omega (-t) \hat{K}(t) dt ~.\]

If we assume that $K(x,y,\gs)=L(x,y)\gd (\gs)$
we get the simpler formula
\[  {\cal L}=\hat{L} ~.\] 
\end{theo}

\section{Exact formulae for  white noises} \label{sec:algeb}

\subsection{Vector valued   white noise}\label{ssec:wn}

Let us assume that $L(x,y)=\gd (x-y){\rm Id}$ meaning that
${\bf f}$ is vector valued white noise on $X\times \R$.
We get, for $\tau >0$,
\[ \Pi =\int _0^{\infty }
\Omega(s) \Omega^\star (s) ds \]
and, assuming $\hat{H}=i\hat{A}+ k $,
with $\hat{A}$ self-adjoint, the following simple formula:
\[  C_{A,B}(\tau )=\frac{1}{2k} P_0(\tau, A,B) ~,\]
which is an exact relation between the correlation and the propagator $P_0$
of  the wave equation without attenuation ($k=0$).

\subsection{Twisted white noise}

This section was motivated by a question of Philippe Roux.
\begin{defi}
A {\rm twisted white noise}  is a random field 
given by ${\bf f}= L_0 {\bf  w}$ with $ {\bf  w}$ a white noise as defined
in Section \ref{ssec:wn} and
 $L_0 \in {\cal L}(\C^N, \C^N)$.
Its  correlation
is 
$ \gd(x-y)\gd (s-s')K_0$ with $K_0= L_0L_0^\star  $.
\end{defi}

We have then 
\[ C_{A,B}(\tau)=[\Omega (\tau)\Pi ](A,B) \]
with 
\[ \Pi =\int _0^\infty \Omega (s)K_0   \Omega^\star  (s)ds ~.\]

In the particular case of the scalar wave equation
with a constant damping $a$ and the dynamics
described in Section \ref{subsection:wave} by Equation
(\ref{equ:wave_1}), we 
get using a gauge transform where
\[ \Omega (t)=e^{-at} \left(
\begin{array}{cc} e^{itP}&0 \\
0& e^{-itP } \end{array} \right) \]
\[ \Pi = \frac{1}{2a} K_{0, {\rm diag }} + R \]
where  $K_{0, {\rm diag }}$ is the diagonal part of $K_0$
and $R$ is going to vanish in the high frequency limit.

\subsection{Wave equations} \label{subsection:wave}
Let us take the case of  a scalar wave equation with constant
damping; the closest results were derived in \cite{WL3,RSK}.

We will consider  
 the  wave equation (\ref{equ:wave})
with 
\begin{itemize}
\item $a>0$ is a  constant damping coefficient
\item $\Delta $ a Riemannian laplacian in some Riemannian manifold $X$,
 possibly
with boundary:
\[ \Delta = g^{ij}(x)\pa _{ij}+ b_i(x)\pa _{i}\]
which is self-adjoint with respect to $|dx|$ and appropriate boundary
conditions; in fact we could replace the Laplacian by any 
self-adjoint operator on $X$!
\item $f=f(x,t)$ the source of the noise which will be assumed to be
a scalar white noise (homogeneous diffuse field):
\[ \E (f(x,s)f(y,s'))=\gd (s-s')\gd (x-y) \]
\end{itemize}

Let us compute the ``causal solution'', i.e. the solution given
by $u={\bf G}f$ with  ${\bf G}$ linear and satisfying
$u(.,t)=0$ if $f(.,s)$ vanishes for $s\leq t $.

We introduce the vector
\[{\bf u}=\left( \begin{array}{c}u\\\pa _t u +au 
\end{array}   \right) \]
which satisfies:
\begin{equation} \label{equ:wave_1}
 \pa _t{\bf u}+a{\bf u} +\hat{H}{\bf u}=\left( \begin{array}{c}0\\
f \end{array}   \right)\end{equation}
with 
\[  \hat{H}     =- \left( \begin{array}{cc}0&{\rm Id}\\ \Delta + a^2&0 
\end{array}   \right) \]

In order to get  a readable expression, it is convenient to introduce
$P=\sqrt{-(\Delta + a^2)}$.
We get then easily:
\[ u(x,t)=\int _0 ^\infty ds \int_X  e^{-as}
\left[ \frac{\sin s P}{P}\right](x,y)f(y,t-s) |dy| \]
where $\sin sP $ is defined from the spectral decomposition of $P$
(any choice of the square root gives the same result for
$\frac{\sin s P}{P}$).
The meaning of the brackets is ``Schwartz kernel of''.
We define 
\[ G_a(t,x,y)=Y(t)\left[e^{-at} \frac{\sin t P}{P}\right](x,y)\]
with $Y$ the Heaviside function. We will call $G_a$ the {\it (causal)
 Green function. } We can rewrite:
\[  u(x,t)=\int_{\R}ds \int_X  
G_a(t-s,x,y)f(y,s) |dy|~. \]

Let us assume now that $f$ is an homogeneous white noise and compute 
the correlation
$C_{A,B}(\tau ):= \lim_{T \ra \infty} \frac{1}{T}\int _0^T
u(A,t)u(B,t-\tau) dt $.

We get quite easily, using
\[ \sin \ga \sin \gb = \ha ( \cos (\ga - \gb) - \cos (\ga + \gb ))~:\] 
\[  C_{A,B}(\tau )=
e^{-a|\tau|}
\left[\frac{\cos \tau P}{4a P^2}
-\frac{a\cos \tau P - P \sin|\tau | P}{4P^2(P^2+A^2)}           \right](A,B) 
\]
or:
\begin{equation}\label{coralg}  C_{A,B}(\tau )=
\frac{e^{-a|\tau|}}{4(P^2+a^2)}
\left[\frac{\cos \tau P}{a}+
\frac{\sin|\tau | P}{P}           \right](A,B) 
\end{equation}

Some comments:
\begin{itemize}
\item {\it Low frequency filtering:} from Equation (\ref{coralg}), we see
for each eigenmode $\Delta u_j= \go _j^2 u_j $ a prefactor, which in
the limit of large $\tau $ is
$\approx e^{|\tau |(r_j -a)} $
for $\go _j ^2=a^2 -r_j^2 < a^2 $. This acts as low frequency  filter
as observed in \cite{RSK}.
\item {\it High frequency regime:} in the high frequency regime,
we get a simplified expression:
\[   C_{A,B}(\tau )\approx 
e^{-a|\tau|}
\frac{\cos \tau P}{4a P^2} \]
of which, taking the derivative w.r. to $\tau $,
we get
\[ \pa _\tau C _{A,B}(\tau )\approx -\frac{ e^{-a|\tau|}}{4a }
\left( \frac{ \sin  \tau P}{P }\right) \]
recovering the expression derived by many authors:
\begin{equation} \label{equ:deriv} 
  \pa _\tau C _{A,B}(\tau )\approx  \frac{ e^{-a|\tau|}}{4a }\left(
 -G(A,B,\tau) 
+G(A,B, -\tau )\right) \end{equation}
\item If the {\it attenuation is small,}
we get 
\[  C_{A,B}(\tau )\approx \frac{1}{4 a}\frac{\cos \tau Q}{Q^2}~, \]
with $Q=\sqrt{-\Delta }$,
and hence the relation (\ref{equ:deriv}).
\end{itemize}

\section{Time reversal symmetry}\label{sec:TRS}

\begin{defi}
\begin{itemize}
\item
A dispersion relation $\D(x,\xi, \go)=0$ 
is said to be time reversal symmetric  (TRS) if it is invariant by
$\ga:(x,\xi)\ra (x,-\xi)$.
\item A linear wave equation (with no attenuation! )
is said to be time reversal symmetric (TRS) if for any solution
${\bf u}(x,t)$ the field $\overline{\bf u}(x,-t) $
is also a solution.
\end{itemize}
\end{defi}
\begin{exem}
\begin{itemize}
\item
Schr\"odinger equations without magnetic fields  
\item Acoustic and elastic wave equations.
\end{itemize}
\end{exem}
\begin{lemm}
If $\D$ is TRS and $\gg(t)=(x(t),\xi(t),\go_0)$ is a
solution of Hamilton's equations, $\ga (\gg(-t))=
(x(-t),-\xi(-t),\go _0)$ too.
\end{lemm}

We have the following:
\begin{prop}
The correlation satifies the following general identity:
\[ C_{A,B}(\tau )=C_{B,A}^\star (-\tau )\]
and, 
in case of a white noise and a time reversible wave dynamics
modified by a constant attenuation (as in Section \ref{ssec:wn}):
\begin{equation}
\label{equ:timereversal} 
 C_{A,B}(-\tau )=\overline{C_{A,B} (\tau )}~.\end{equation}
\end{prop}
Approximations of Equation (\ref{equ:timereversal}) turn out to 
be important in applications to clocks synchronisation.

\section{Random fields and pseudo-differential operators }\label{sec:random}
\subsection{Goal}
Our aim in this section is to build quite general random fields 
with  correlation distances given by a small parameter $\ge$.
It seems to be  natural for that purpose to use $\ge-$pseudo-differential
operators. We will see how to compute the generalized power 
spectrum using Wigner measures.

\subsection{White noises}

Let $({\cal H},\langle .|.\rangle )$ be an Hilbert space.
 There exists a canonical {\it Gaussian random
field} on it,  called the {\it white noise} and denoted by
${\bf w}_{\cal H}$ (or simply ${\bf w}$ if there is no possible confusion).
This random field is defined by the properties that:
\begin{itemize}
\item For all $\vec{e}\in {\cal H}$,
\[ \E (\langle  {\bf w}|\vec{e} \rangle)=0 \]
\item 
For all $\vec{e},\vec{f} \in {\cal H}$,
\[ \E ( \langle  {\bf w}|\vec{e} \rangle\overline{\langle  {\bf w}|\vec{f} \rangle})
=\langle  \vec{e}|\vec{f} \rangle\]
\end{itemize}

Unfortunately, ${\bf w}$ is not a random vector in ${\cal H}$ unless
  $\dim {\cal H} <\infty $\footnote{If ${\bf w}$
 were a vector in ${\cal H}$,
we would have ${\bf w}= \sum \langle {\bf w}|\vec{e_j}
 \rangle\vec{e_j} $ for any orthonormal 
basis $(\vec{e_j })$ and we see
that 
\[ \E (\| {\bf w}\| ^2)=
 \sum \E (\langle {\bf w}|\vec{e_j} \rangle)^2 =\dim {\cal H}~.\]} ,
but only a random {\it Schwartz distribution}.

We have nevertheless the following usefull proposition:
\begin{prop}
If $A $ is an Hilbert-Schmidt operator on ${\cal H}$,
the random field $A{\bf w}$ is almost surely in ${\cal H}$.
\end{prop}
\begin{demo}
$\E (\langle A{\bf w}|A{\bf w} \rangle)=\E (\langle  A^\star A{\bf w}|{\bf w} \rangle)= {\rm Trace }(A^\star A )$
which is finite, by definition, exactly for Hilbert-Schmidt operators.
\end{demo}
\subsection{Examples}
\begin{exem} {\rm Stationary noise on the real line:}
let us take a random field on the real line which is given
by the convolution product of the scalar white noise $w$
with a fixed smooth compactly supported function $F$:
$f=F \star { w} $.
 Then $f$ is stationary: it
means that the correlation kernel $K(t,t')=\E (f(t)\overline{f(t')})$
is a fonction of $t-t'$.
On the level of Fourier transforms
$\hat{f}=\hat{F}\hat{{ w}} $, 
$\E(\hat{f}(\go)\overline{\hat{f}(\go')})= |\hat{F}|^2(\go)\gd (\go -\go')$
and the positive function $|\hat{F}|^2(\go)$ is usually called 
the {\rm power spectrum} of the stationary noise.

\end{exem}
\begin{exem} If $X$ is a $d$-dimensional bounded domain.
Let us denote  the Sobolev spaces on $X$ by $H^s(X)$.
If $P: L^2(X) \ra H^s(X) $ with $s > d/2$, $P{ w}$ is in $L^2(X)$.
\end{exem}

\begin{exem} {\rm Brownian motions:}
if $X=\R$, ${ w}$ is the {\it derivative} of the
Brownian motion: if $b(t)=\int _0^t { w}(s)ds $,
$b:[0,+\infty [ \ra \R $ is the Brownian motion which
is in $L^2 ([0,T])$ for all finite $T$.
\end{exem}

\begin{exem} If $X$ is  a smooth compact manifold or domain
and $P$ is smoothing, meaning that $P$ is given by an integral
smooth kernel
\[ Pf(x)=\int _X [P](x,y) f(y) |dy | ~,\]
$F= P{ w} $ is a random smooth function. Its 
 correlation kernel
\[C(x,y):= \E (F(x)\overline{ F(y)})\]
is given by:
\[  [P P^\star](x,y)=
\int _X [P](x,z)\overline{[P](y,z)} |dz| ~.\] 

\end{exem}

\begin {exem}{\rm Random vector fields:}
let us consider ${\cal H}=L^2(X,\R ^N)$. For example,
in the case of elasticity, $X$ is a 3D domain and $N=3$. The field here
 are just fields of infinitesimal deformations (a vector field).

\end{exem}

\subsection{Modelling the noise using pseudo-differential
 operators}\label{sec:noise}

The main goal of the present section  is  to build natural 
random fields 
which are  non homogeneous  with small distances
of correlation of the order of $\ge \ra 0$.
The noise is non homogeneous in $X$, but could also
be non  isotropic w.r. to directions.

\subsubsection{Noises from pseudo-differential operators}

 It is therefore natural to take for
noise on a manifold $Z$ 
the image of an homogeneous white noise by a pseudo-differential operator
${N}$ of smooth compactly supported 
symbol $n(z,  \zeta)$.
The correlation $C(z,z')$ will then be given 
as the Schwartz kernel of $N N^\star $
which is a  $\Psi DO$ of principal symbol $|n|^2 $.
We have
\[ C(z,z') \sim \ge ^{-d}\tilde{|n|^2 }\left(z, \frac{z'-z}{\ge }\right)~,\]
while $|n|^2 (z, \zeta  ) $ is the ``power spectrum'' of the noise
at the point $z$.

This construction gives smooth random fields which can be localized 
in some very small domains of the manifold $Z$,
which are non isotropic and
which have small distance of correlations.
Moreover it will allow to use technics of 
microlocal analysis with the small parameter given by
$\ge $. 

\subsubsection{Power spectrum and Wigner measures}

\begin{defi}
If $f=(f_\ge) $ is a suitable family of functions on $Z$,
 the {\rm  Wigner measures} $W_f^\epsilon$ of $f$
  are the signed measures on the
phase space $T^\star Z$ defined by
\[ \int a dW_f^\epsilon := \langle  {\rm Op}_\ge (a) f_\ge|f_\ge  \rangle ~. \]
The measures $dW_f^\ge $ are the phase space {\rm densities  of energy}
of the functions $f_\ge $.
\end{defi}

We  now define: 
\begin{defi}  The {\rm power spectrum} of the  random field 
$f=(f_\ge)$ is the phase space density  $P_f^\ge $ defined by:
\[ P_f^\ge = \E (W_f^\ge ) ~:\]
the power spectrum of a random field
 is  its  average Wigner measure.
\end{defi}

\begin{prop}
 The power spectrum $P$
of  $f_\ge= {\rm Op}_\ge (n) {w}$,
satisfies:
\[ P_f^\ge \sim (2\pi \ge)^{-d}|n|^2(x,\xi)|dx d\xi | ~.\]
\end{prop}
\begin{demo} Let us put $N={\rm Op}_\ge (n)$, we have
\[ \langle  {\rm Op}_\ge (a)N{w}|N{w} \rangle=
 \langle  N^\star {\rm Op}_\ge (a)N{w}|{w} \rangle \]
and $\E (\langle  A{w}|{w} \rangle )={\rm trace}(A)$. We get
\[ \E (\int a dW _f ^\ge) =
{\rm trace}(N^\star {\rm Op}_\ge (a) N) \]
which can be evaluated using the $\Psi$DO calculus
as
\[  \E (\int a dW _f ^\ge)\sim (2\pi \ge)^{-d}\int a|n|^2 dxd\xi ~.\]

\end{demo}

\subsubsection{Space-time noises}
If  $Z=X\times \R $ is the space-time,  we will
take our noise as before $f=L{w}$; we will assume
 the noise {\it homogeneous in time,}
the symbol $l$ of $L$  is assumed to be given
by $l( x,\xi,\go)$.

In this case, the correlation is given by:
\begin{equation} \label{equ:psido}
 K(x,y; t)=[ L L^\star](x,y;0,t)\end{equation}
which is the  Schwartz kernel of a  $\Psi DO$ of \ppal symbol
$l l^\star  (x,  \xi;\go )$.

\subsection{Equipartition and polarizations}

If ${\bf f}_\ge: Z\ra \R^q $ is a vector valued family of
functions, the Wigner measure is matrix valued:
if ${\bf a}:T^\star X \ra {\rm Sym}(\R ^q)$,
we define
\[ \int {\bf a}dW_{\bf f}^\ge :=
\langle  {\rm Op}_\ge ({\bf a}){\bf f}_\ge|{\bf f}_\ge \rangle \]
If we have, for each point $(z,\zeta)$ of $T^\star Z$ a splitting 
$\R^q =\oplus E_j (z,\zeta) $ with projectors $P_j$, there exists 
 canonical measures $ \mu _j $ defined
by
\[ \int {\bf a} d \mu _j ={\rm Trace~}({\bf a}P_j )|dzd\zeta |~. \] 

If the $E_j$'s are defined by the
polarizations of an Hamiltonian $\hat{H}$,
the {\it microcanonical Liouville measures} are 
given by
\[ \int {\bf a}dL _E :=\sum _j \int _{\gl _j \leq E}  {\bf a}
{d\mu _j}
~. \]

A (random) state of energy $E$ of $\hat{H}$ is said to equipartited
if its (average) Wigner measure converges to the microcanonical measure.
If $P_E$ is the spectral projector of $\hat{H}$ over the modes
of energies less than $E$, the random state $P_E w$ is equipartited.

\section{High frequency limit
 of the correlation: Schr\"odinger equations}\label{sec:pseudos}

The main result is easier to derive in the case of a scalar field 
gouverned by a wave equation which gives the first order
derivative of the field: it is a generalization of the Schr\"odinger equation.

\subsection{Assumptions}\label{subsec:assumptions}
Let us start with the semi-classical Schr\"odinger like  equation
\[ \frac{\ge}{i}{u}_t +  \hat{H}_\ge{u}=
 \frac{\ge}{i}{ f} \]
where
\begin{itemize}
\item $ \hat{H}_\ge = \frac{\ge }{i} \hat{H}$
is an {\bf $\ge-$pseudo-differential operator:}
\[   \hat{H}_\ge :={\rm Op}_{\ge}
(H_0+\ge H_1) \]
with  
\begin{itemize}
\item The {\bf principal symbol} $H_0(x,\xi):T^\star X\ra  \R $
which gives the classical (``rays'') dynamics:
\[ \frac{dx_j}{dt}=\frac{\pa H_0}{\pa \xi_j},~
 \frac{d\xi_j}{dt}=-\frac{\pa H_0}{\pa x_j},~1\leq j \leq d~.\]
We will denote by $\phi _t$ the flow of the previous vector
field.
\item 
For technical reasons (see Lemma \ref{lemm:funct}),
 we assume that $H_0$ is {\bf elliptic at infinity:}
$\lim _{\xi \ra  \infty }H_0(x,\xi)=+\infty $. 
We define $\hat{H}_0= {\rm Op}_\ge (H_0)$ and 
the unitary group $U(t)={\rm exp}(-it\hat{H}_0/\ge )$.
We define also $\hat{H}_1 ={\rm Op}_\ge (H_1)$.
\item The {\bf sub-principal symbol} $H_1(x,\xi)$ admits some
positivity property which controles the attenuation:
there exists $k>0$, such that
\[ \Im H_1 \leq -k ~.\] 
\end{itemize}
\item The {\bf random field}
 $f$ is given by ${ f} ={\rm Op}_\ge (l(x,\xi,\go))w$
with $w$ the white noise on $X\times \R$ and 
with $l$ smooth, compactly supported
w.r. to $(x,\xi) $ and whose Fourier transform w.r. to
$\go $ is compactly supported. The power spectrum of ${ f}$
is $(2\pi \ge)^{-(d+1)}|l|^2  (x,\xi,\go)$.
\end{itemize}

The previous assumptions will be used everywhere inside 
Section \ref{sec:pseudos}. 

\subsection{Subprincipal symbols and attenuation}

\begin{lemm} \label{lemm:subp}
Under the assumptions of Section \ref{subsec:assumptions},
we have:

a) There exists $c>0$ so that, for $|t| \leq c |\log (\ge)|$,
$ \Omega (t)=U(t) Y(t) $
with $Y(t)$ a \OPD~ of principal symbol
${\rm exp }(-i \int  _0 ^t H_1 (\phi _s (x, \xi)) ds )$.

b)  for all $t\geq 0$,
 the estimates
\[ \| Y(t) \| =O(e^{-k't })\]
with $\|. \| $ the operator norm in $L^2(X)$
and $0<k'<k$.
\end{lemm}
\begin{demo}
a) We start with $\Omega (t)=U(t)Y(t)$ and hence
\[ Y'(t)+iU(-t)\hat{H}_1 U(t)Y(t)=0 ~.\]
Using Egorov's Theorem with logarithmic  times,
 as in Section \ref{sec:egorov},
we get
\[ Y'(t)+ i\hat{H}_1(t) Y(t) =0 \]
with $\hat{H}_1(t)$ a \OPD~ of principal symbol
$iH_1 (\phi _t (x,\xi))$. It is then enough to start with a formal
expansion in $\ge$ of the symbol of $Y(t)$ and to work
by induction on the powers of $\ge $.

b) We have
\[ \frac{ d}{dt} \langle v(t)|v(t) \rangle =
2 \Re \langle v(t)| -i\hat{H}_1(t)v(t) \rangle \]
and we use G{\aa}rding inequality (see \cite{Di-Sj}): if $a\geq 0$,
${\rm Op }_\ge (a) \geq -C $
for any $C>0$ and $\ge $ small enough.
\end{demo}

\subsection{Some lemmas}
\begin{lemm}\label{lemm:funct}
If $A={\rm Op}_\ge (a)$ with $a$ compactly supported,
the operator
$B={\rm exp}(it\hat{H}) A $ is a \OPD ~
of principal symbol
$b={\rm exp}(itH_0) a $.
\end{lemm}
\begin{demo}
Let us choose a function $\chi \in C_o^\infty (\R, \R)$ 
so that $\chi (H_0)$ is equal to $1$ in some neighbourhood 
of the support of $a$. We have
\[ B={\rm exp}(it\hat{H}_0)\left(\chi (\hat{H}_0)+
(1-\chi (\hat{H}_0))\right)C \]
with $C=Y(-\ge t)A $ is a compactly supported \OPD~ of 
principal symbol $a$.
The previous expression of $B$ splits into 2 terms $B=I+II$.
The first one rewrites
$I=\Phi (\hat{H}_0) C $ with $\Phi=e^{it.}\chi \in C_o^\infty $
which is a \OPD~ of principal symbol $e^{itH_0}a $
thanks to the functional calculus of elliptic self-adjoint \OPD's
(see \cite{Di-Sj},  Chap. 8).
The second one is smoothing.
\end{demo}

The following Lemma follows directly from the definitions:
\begin{lemm} \label{lemm:symbol}
If  ${K}(x,y,t )$ is the correlation kernel of $f$, then
   $\hat{K}(\gs_1 )$, 
the operator whose kernel is $K(.,.,\ge \gs_1 )$,
 is a $\Psi$DO which vanishes for $|\gs_1 |\geq C  $
and whose principal symbol is $(2\pi \ge)^{-1}\int e^{i\gs _1 \go }
|l|^2(x,\xi,\go) d\go $.
\end{lemm}

From Equation (\ref{equ:correl}), we get, for $\tau >0 $
and $\ge \leq C/\tau $:
\begin{equation} \label{equ:L}
 C_{A,B}(\tau)=
[\Omega (\tau) \int _0^\infty ds \Omega (s)
 {\cal L} \Omega ^\star (s)  ](A,B) \end{equation}
with 
\[  {\cal L} = \ge 
\int _{|\gs _1|\leq C  } \Omega (-\ge \gs_1) \hat{K}(\gs _1)d\gs_1 ~.\]

\begin{lemm}\label{lemm:psido}
${\cal L} $ is a $\Psi$DO of principal symbol
$|l|^2(x,\xi,-H_0 (x,\xi))$.
\end{lemm}
\begin{demo}
The result follows from Lemma \ref{lemm:funct} 
and the value of the symbol of $\hat{K}(\gs_1)$ 
given in Lemma \ref{lemm:symbol}, by integrating
w.r. to $\gs$.
\end{demo}

\subsection{Applying Egorov Theorem}\label{sec:egorov}

We can apply Egorov Theorem:
\begin{theo} \label{theo:egorov}  {\bf (Egorov's Theorem)}
If $A={\rm Op}_\ge (a)$,
$U(-t)A U(t)=A_t$ where $A_t $ is a ~\OPD~ of principal symbol
$a\circ \phi _t $ with  $\phi _t$ is the Hamiltonian flow of $H_0$.
\end{theo}
We will need a {\it large  time}
 estimation in the Egorov's Theorem. Such estimates
are provided in the nice paper \cite{Bo-Ro}:
Egorov's Theorem still works under suitable hypothesis on $H_0$
for time bounded by $c|\log \ge |$ where $c$ is related to the 
Liapounov exponent of the classical dynamics.
Such time is called {\it Eherenfest time} and will be denoted by 
$T_{\rm Ehrenfest}$.

We
get the main result: 
\begin{theo}\label{theo:N=1}
With the assumptions of Section \ref{subsec:assumptions},   the correlation
 is given, for $\tau >0$,  by 
\[ C_{A,B}(\tau)=[\Omega (\tau)\circ  \Pi ) ](A,B)\]
where $\Pi={\rm Op}_\ge (\pi)
+R $
with:
\[ \pi (x,\xi)=\int _{-c |\log \ge |}^0  
{\rm exp}\left(2 \int _{t}^0 \Im  (H_1)
 (\phi_s (x,\xi))ds \right)|l|^2 
\left(\phi_t (x,\xi),-H_0(x,\xi)\right) dt ~.\]
and $R$ the remainder term 
is ``$O(\ge ^\ga )$''.
More precisely, let us consider $C_{A,B}(\tau )$ as the Schwartz kernel
of an operator $\hat{C}(\tau )$. This operator is Hilbert-Schmidt 
\footnote{An Hilbert-Schmidt operator $A$ is an operator
whose Schwartz kernel $[A](x,y)$ is in $L^2(X\times X)$ and the
Hilbert-Schmidt norm  $\| A  \| _{\rm H-S}$ of $A$ 
is the $L^2$ norm of $[A]$.}                 
with an
Hilbert-Schmidt norm of the order of $\ge ^{-d/2}$. We have 
\[ \| \hat{R} \|_{\rm H-S}=O(\ge^{\ga - d/2 })~.\] 
\end{theo}

\begin{demo}
We start from 
\begin{equation} \label{equ:integral}
 \Pi= \int _0^\infty \Omega (s) {\cal L} \Omega ^\star (s) ds \end{equation}
as given by Equations (\ref{equ:correl}) and (\ref{equ:L}).
We have  $\Omega (s)=U(s)Y(s) $
with  $Y (s) $ is a ~\OPD ~ of principal symbol
${\rm exp}({-i\int _0^s   H_1 (\phi_u (x,\xi))du}) $ and
\[\Omega (s) {\cal L}   \Omega^\star  (s)=
U(s)\left( Y(s) {\cal L} Y^\star (s) \right) U (-s)\]
to which we want to  apply Egorov's Theorem.
There is a technical problem due to the fact that Egorov's Theorem
is only valid untill  Ehrenfest times.
We split the integral (\ref{equ:integral}) into two parts
$\Pi = \int _0 ^{T_{\rm Ehrenfest}}+ \int _{T_{\rm Ehrenfest}}^\infty $.
The first term is estimated using Egorov's Theorem for large times. An upper
bound for second part follows from the decay estimate of 
$\| \Omega (t)\| $ given in Lemma \ref{lemm:subp} and the estimates
$\| {\rm Op}_\ge (a) \| _{\rm H-S}=O(\ge ^{-d/2})$.


\end{demo}

Assuming still  $\tau >0$,
 we see that the correlation $ C_{A,B}(\tau)$ 
 is close to the kernel
of a Fourier integral operator associated to the canonical
transformation $\phi _\tau $. It is given as a sum over all  classical
trajectories $\gg $ from $B$ to $A$ in time $\tau$
of Cauchy data $(B,\xi_B)$ with $H_0(B,\xi_B)=-\go $   for which
the backward trajectories crosses the support of 
the power spectrum $ll^\star  (.,.,\go )$.
If $\gg $  is such a trajectory and $B$ and $A$ are non conjugated along it,
this contribution is given by the well known 
Van Vleck formula\footnote{The Van Vleck formula expresses the propagator
$ P(\tau ,A,B) $ as a sum of $p_\gg =
(2\pi i\ge )^{-d/2}a_\gg (\ge){\rm exp}(iS(\gg)/\ge )$
with $a_\gg (\ge) $ a formal power series in $\ge$ with a first term
explicitely computable}
multiplied by $\pi (B,\xi_B)$.

\begin{coro} Let $K$ be the support of 
$l(x,\xi,-H_0(x,\xi))$ and $K_\infty $ the smallest closed set of $T^\star X$
invariant by the Hamiltonian flow of $H_0$ and containing $K$.
The Hamiltonian $H_0$ restricted to $K_\infty $ can be recovered
from the knowledge of $\hat{C}(\tau )$ for
$0<|\tau |\leq \tau _0 $.

In particular, if there exists $(x, \xi)$ with $H_0(x,\xi)=E$
and $l(x,\xi,-E)\ne 0 $ and if  $\phi _t$ is ergodic on
$H_0^{-1}(E)$,  then we can recover the flow $\phi _t$ on
$H_0^{-1}(E)$.
\end{coro}

\subsection{Remarks}

We would like to extend the previous approach to 
the general case, i.e. to $N>1$.
There are 2 difficulties to overcome:
\begin{itemize}
\item One has to extend Lemma \ref{lemm:psido} to the case of systems
\item Egorov Theorem is no  more true, but remains true 
on average as in Lemma \ref{lemm:egorov_aver}
\end{itemize}

\section{High frequency limit of the correlation: wave equations}
\label{sec:wave}

We want to derive results similar to those of Theorem \ref{theo:N=1}
in the case of the scalar wave equation ($N=2$)
given as follows:
\[ u_{tt}+ 2au_t -\Delta u= f \]
where $a>0$ is  {\it constant}  and $-\Delta $ is the Laplace-Beltrami operator
on a smooth complete Riemannian manifold $X$.

We will assume that the source noise $f$ is given by
$f=Lw $ with $w $ a white noise on $X \times   \R $
and $L={\rm Op}_\ge (l)$ with $l=l(x,\xi)$ smooth, with compact support, 
and independant of $\go$. We will moreover assume time reversal symmetry of the
noise, namely $l(x,-\xi)= l(x,\xi)$; it implies that the kernel
$[L](x,y)$ is real valued.

\subsection{Direct derivation}

Let us introduce $P=\sqrt{-\Delta -a^2}$, the causal solution 
of Equation (\ref{equ:wave}) is given, as already used in Section \ref{subsection:wave}, 
by 
\[ u(x,t)=\int _0 ^\infty ds \int _X G(s,x,y)u(t-s,y) |dy| \]
with the Green function 
\[ G(t,x,y)=Y(t)e^{-at}\frac{\sin tP}{P} ~.\]
By a direct  calculation and denoting by $\hat{K}=P^{-1}
LL^\star P^{-1}$ with $L={\rm Op }_\ge (l)$,
we get, for $\tau >0$,
\begin{equation} 
\label{equ:correl_wave}
 C_{A,B}(\tau) =
\ha e^{-a\tau }\left[
\Re \left( 
e^{-i\tau P}\int _{-\infty}^0 e^{2as}e^{isP}\hat{K}
\left(e^{-isP}-e^{isP}\right)ds
\right)\right](A,B) ~.
\end{equation} 

This integral splits into 2 parts, the first one can be asymptotically
computed using  Egorov's Theorem as in the proof of Theorem
\ref{theo:N=1}, while the second is
smaller  by the
\begin{lemm}\label{lemm:egorov_aver}
If $\hat{A}$ is a  compactly supported ~\OPD~ of order $0$ and if
\[ J:=\int _{-\infty}^0 e^{2as}e^{isP}\hat{A}
e^{isP} ds ~,\]
we have 
\[ \| J \| _{\rm H-S}= O(\ge ^{\gg -\frac{d}{2}}) ~,\]
with some non negative $\gg$.
\end{lemm}
This is proved by cuting the integral into 2 pieces
as in the proof of Theorem \ref{theo:N=1},
 using Egorov's Theorem  and  integrating  by part.

The final result is:
\begin{theo}
With the previous assumptions, 
we have, for $\tau >0$:
\[ C_{A,B}(\tau) =
\frac{\ge ^2}{2}  e^{-a\tau}\left[ \cos \tau P \circ \Pi \right] (A,B)
+R  \]
with $\Pi $ a ~\OPD~ of principal symbol
\[ \pi (x,\xi)= \| \xi \| ^{-2}
\int _{-\infty }^0 e^{2as }|l|^2(\phi _s (x,\xi)) ds ~,\]
and $\| R \| _{\rm H-S}=O(\ge ^{\gg +2 -d/2})$ with $\gg >0$. 
\end{theo}
\begin{rem}
The prefactor $\ge ^2$ is just here because we have to pass
from the ~\OPD's without small parameter $P$ to an $\ge-$\OPD:
 $ P=\ge^{-1}{\rm Op}_\ge (\| \xi \|) +l.o.t.$.
\end{rem}

\subsection{Using the general formalism}
We can again start from Equation (\ref{equ:Pi}). We are reduced to calculate
the correlation between $u(A,t)$ and $u(B,t)$ which is given by 
\[ C_{A,B}^{11}(\tau)=[ \Omega (\tau)\Pi )]^{11}(A,B)~.\]

We should first put the wave equation as a first order semi-clasical
equation. We put
\[ {\bf u}=\left( \begin{array}{c} u \\\ge (u_t+au )
\end{array} \right) ~.\]
and
\[-i \ge \hat{H}=
\left( \begin{array}{cc} -ia &1 \\
i\ge ^2 (\Delta +a^2)& -ia 
\end{array} \right) ~.\]
We get, by definig $P=\sqrt{-(\Delta + a^2)}$,
\[ \Omega (t)=e^{-at} \left( \begin{array}{cc} \cos tP &\frac{\sin tP}{\ge P}
 \\-\ge P \sin tP& \cos tP 
\end{array} \right) ~.\]
Moreover, if $f={\rm Op}_\ge (m)w $,
\[ {\bf f}=\left( \begin{array}{c} f \\0 \end{array} \right) 
 = {\rm Op}_\ge (l){w} \]
with 
\[ l= \left( \begin{array}{cc}m  &0\\ 
0& 0 \end{array} \right) ~.\]

We see that the computation is less easy $\cdots $

\section{What's left?} \label{sec:problems}

There are still several problem to discuss:
\begin{itemize}
\item The case of vector valued wave equations with several
polarizations.
\item The precise study of autocorrelations:
as mentionned to me by U. Smilanski, the autocorrelations can
be usefull in order to learn more about the source
of the noise ${\bf f}$.
\item The case of surface waves: effective Hamiltonains
and the associated inverse spectral problems will be discussed
in \cite{YCdV2}.
\item The case of the source noise located on the boundary:
\[ \left\{ \begin{array}{l}
u_{tt}+ au_t -\Delta u =0\\
u_{\pa X}=f \end{array}\right. \]
will be discussed
in \cite{YCdV3}.

\end{itemize}

\section{Random scattered waves}\label{sec:scattering}

In this last independant section,
 we will revisit what was maybe the starting point
of this story by Keiiti Aki in the fifties: he wanted to measure
the speed of propagation  of seismic plane waves
by averaging over the incidence directions. It turns out that
we get nice formulae even for non homogoeneous media.

\subsection{Introduction}

Let us consider the propagation of waves outside a compact domain $D$ in
the Euclidian space $\R^d $.
Let us put $\Omega = \R^d \setminus D $.
 We can assume for example
Neumann boundary conditions. We will denote by $\Delta _\Omega $
the previous self-adjoint operator.
So our stationary wave equation  is the Helmoltz equation
$\Delta _\Omega f + k^2 f=0 $ with the boundary conditions.
We consider a bounded intervall $I=[E_-, E_+]\subset ]0,+\infty [$
and the Hilbert subspace ${\cal H}_I $ of $L^2(\Omega )$  which is the image 
of the spectral projector $P_I $ of our Laplace operator  $\Delta _\Omega $.

Let us compute the integral kernel  $\Pi_I (x,y)$  of $P_I$
defined by:
\[ P_I f(x)=\int _\Omega \Pi _I (x,y) f(y) |dy| \] 
 into 2 different ways:
\begin{enumerate}
\item From general spectral theory
\item From  scattering theory.
\end{enumerate}
Taking the derivatives of $\Pi_I (x,y)$  w.r. to $E_+$, we get
a simple general and exact relation
between the correlation of scattered waves and the Green's function
confirming the calculations from \cite{SS}
in the case where $D$ is a disk. 

\subsection{$\Pi_I(x,y)$             from spectral theory}

Using the resolvent kernel (Green's function)
$G(k,x,y)=[ (k^2  + \Delta_\Omega ) ^{-1 } ](x,y) $
for $\Im k >0 $
and the Stone formula, we have:
\[ \Pi_I(x,y)=\frac{2}{\pi }\Im \left(\int _{k_-}^{k_+}
G(k+i0, x,y) kdk \right)  \] 
Taking the derivative w.r. to $k_+$ of $\Pi_{[E_-,k^2]}(x,y)$,
we get 
\begin{equation} \label{equ:stone}
 \frac{d}{dk} \Pi_{[E_-,k^2]} (x,y)=\frac{2 k}{\pi }\Im (
G(k+i0, x,y))  ~. \end{equation}

\subsection{Short review of scattering theory}

They are many references for scattering theory: for example \cite{RS3}.

Let us define  the plane waves
\[ e_0 (x,{\bf k})=e^{i<{\bf k}|x> }~.\]
We are looking for solutions 
\[ e (x,{\bf k})= e_0 (x,{\bf k})+ e^s(x,{\bf k})\]
of the 
Helmoltz equation
 in $\Omega $
 where $e^s$, the scattered wave satisfies the so-called
Sommerfeld radiation condition:
\[  e^s(x,{\bf k})=\frac{e^{ik|x|}}{|x|^{(d-1)/2}}\left( e^{\infty}
(\frac{x}{|x|},{\bf k }) + O(\frac{1}{|x|}) \right),
~x\ra \infty ~.\]
The complex function $e^{\infty}
(\hat{x},{\bf k })$ is usually called the 
{\it scattering amplitude}.

It is known  that the previous problem admits an unique solution.
In more physical terms, $e(x,{\bf k})$ is the wave generated by the full
scattering process from the  plane wave $e_0 (x,{\bf k})$. 
Moreover we have a generalized Fourier transform:
\[ f(x)=(2\pi )^{-d}\int_{\R ^d} \hat{f}({\bf k }) e (x,{\bf k})|d{\bf k}|\]
with 
\[ \hat{f}({\bf k })=\int_{\R ^d}  \overline{e (y,{\bf k})}f(y) |dy|~. \]

From the previous generalized Fourier transform, we can get the
kernel of any function $\Phi (-\Delta _\Omega )$ as follows:
\begin{equation} \label{equ:function}
 [\Phi (-\Delta _\Omega )](x,y)=
(2\pi )^{-d}\int_{\R ^d}\Phi (k^2) e (x,{\bf k})\overline{e (y,{\bf k})}
|d{\bf k}|~.\end{equation}

\subsection{$\Pi_I(x,y)$    from scattering theory}

Using  Equation (\ref{equ:function}) with
$\Phi =1_I $ the 
 characteristic functions
of some bounded intervall $I$,
  we get:
\[ \Pi_I (x,y)=
(2\pi)^{-d}\int _{E_- \leq {\bf k}^2 \leq E_+ } e(x,{\bf k})
\overline{e(y,{\bf k})} |d {\bf k}| ~.\]
Using polar coordinates and defining $| d\gs |$
as the usual measure on the unit $(d-1)-$dimensional sphere, we get:
\[ \Pi_I (x,y)=(2\pi)^{-d}\int _{E_- \leq k^2 \leq E_+ }\int _{{\bf k}^2=E}
 e(x,{\bf k})
\overline{e(y,{\bf k})}k^{d-1}dk | d\gs | ~.\]
We will denote by $\gs _{d-1}$ the total  volume
 of the unit sphere in $\R^d$: $\gs _0=2,~\gs _1=2\pi,~\gs _2= 4\pi,\cdots $.

Taking the same derivative as before, we get:
\[  \frac{d}{dk} \Pi_{[E_-,k^2]}
  (x,y)=\frac{k^{d-1}}{(2\pi)^d}\int _{{\bf k}^2=E}
 e(x,{\bf k})
\overline{e(y,{\bf k})} |d\gs| ~.\] 

This integral can be interpreted, using 
the correlation $C_E^{\rm scatt}(x,y)$
 of random scattered waves 
of energy $E$ defined by
\[  C_E^{\rm scatt}(x,y)=\frac{1}{\gs _{d-1}}\int _{{\bf k}^2=E}
 e(x,{\bf k})
\overline{e(y,{\bf k})} |d\gs |, \]
as 
\begin{equation}  \label{equ:eigen}
 \frac{d}{dk} \Pi_{[E_-,k^2]} (x,y)=\frac{k^{d-1}\gs_{d-1}}{(2\pi)^d}
  C_E^{\rm scatt}(x,y)~. \end{equation}

\subsection{Correlation of scattered plane waves and Green's function:
the scalar case}

From Equations (\ref{equ:stone}) and (\ref{equ:eigen}), we get:
\[ \frac{k^{d-1}\gs _{d-1}}{(2\pi)^d} C_E^{\rm scatt}(x,y)=
                    \frac{2k}{\pi }\Im (
G(k+i0, x,y))  ~. \]
Hence
\[  C_E^{\rm scatt}(x,y)=\frac{2^{d+1} \pi ^{d-1}  }{k^{d-2}\gs _{d-1}  }
\Im (G(k+i0,x,y))~. \]
For later use, we put
\begin{equation}
\label{equ:gamma}
 \gg_d(k)=\frac{2^{d+1} \pi ^{d-1}  }{k^{d-2} \gs _{d-1}}~.
\end{equation}

\subsection{The case of elastic waves}

We will consider the vectorial stationary elastic wave equation
in the domain $\Omega $:
\[ \hat{H}{\bf u} - \go ^2  {\bf u} =0, \]
with symmetric boundary conditions,
where 
\[ \hat{H}{\bf u}=-a ~\Delta {\bf u} - b ~{\rm grad}~{\rm div }{\bf u}~. \]
where $a$ and $b$ are constant:
\[ a= \frac{\mu}{\rho},~b=\frac{\gl + \mu}{\rho}\]
with $\gl,~\mu $ the Lam\'e's coefficients and $\rho $
the density of the medium. 

\begin{itemize}
\item {\it The case $\Omega = \R ^d $}

We want to  derive the spectral decomposition of $\hat{H}$ from
the Fourier inversion formula. Let us choose,  for ${\bf k}\ne 0 $,
by $\hat{\bf k}, \hat{\bf k}_1, \cdots, \hat{\bf k}_{d-1}  $
 an orthonormal basis of  $\R ^d $ with $\hat{\bf k}=\frac{\bf k}{k}$
such that these vectors depends in a measurable way of ${\bf k}$.
Let us introduce 
$P_P^{\bf k}=\hat{\bf k}\hat{\bf k}^\star $ the orthogonal projector
onto $\hat{\bf k}$ and $P_S^{\bf k}
=\sum_{j=1}^{d-1} \hat{\bf k}_j\hat{\bf k}_j^\star $
so that $P_P + P_S= {\rm Id}$.
Those projectors correspond respectively to the polarizations
of $P-$ and $S-$waves. 

We have 
\[\begin{array}{l}
 \Pi _I (x,y)=
(2\pi)^{-d}
\int _{\go ^2 \in I} \go ^{d-1} d\go \left( 
(a+b)^{-d/2}\int _{k^2=\go ^2/(a+b)^2}e^{i{\bf k}(x-y)}P_P^{\bf k} d\gs
+\right. \\
\left. a^{-d/2}\int _{k^2=\go ^2/a^2}e^{i{\bf k}(x-y)}
P_S^{\bf k} d\gs \right) ~. \end{array} \]       
using the plane waves
\[ e_P^O (x,{\bf k})=e^{i{\bf k}x} \hat{\bf k} \]
and 
\[ e_{S,j}^O (x,{\bf k})=e^{i{\bf k}x} \hat{\bf k}_j \]
we get the formula:
\[ \begin{array}{l} \Pi _I (x,y)=
(2\pi)^{-d}
\int _{\go ^2 \in I} \go ^{d-1} d\go \left( 
(a+b)^{-d/2}\int _{k^2=\go ^2/(a+b)^2}  e_P^O (x,{\bf k})
 (e_P^O (y,{\bf k}))^\star  d\gs
+\right.\\\left.
a^{-d/2}\sum_{j=1}^{d-1} \int _{k^2=\go ^2/a^2}
  e_{S,j}^O (x,{\bf k}) (e_{S,j}^O (y,{\bf k}))^\star
 d\gs \right) ~. \end{array}       \]

\item {\it Scattered plane waves}

There exists scattered plane waves
\[  e_P (x,{\bf k})= e_P^O (x,{\bf k})+ e_P^s (x,{\bf k})\]
\[ e_{S,j} (x,{\bf k})= e_{S,j}^O (x,{\bf k})+ e_{S,j}^s (x,{\bf k})\]
satisfying the Sommerfeld condition and from which we can deduce
the spectral decomposition of $\hat{H}$.

\item {\it Correlations of scattered plane waves and Green's function}

Following the same path as for scalar waves, we get an 
 identity which holds now for the full Green's tensor
$\Im {\bf G} (\go +iO, x,y)$:
\[\begin{array}{l} \Im {\bf G} (\go +iO, x,y)=\gg_d (\go) \left(
(a+b)^{-d/2}\int _{k^2=\go ^2/(a+b)^2}  e_P (x,{\bf k})
 (e_P (y,{\bf k}))^\star  d\gs\right.
+\\\left. 
a^{-d/2}\sum_{j=1}^{d-1} \int _{k^2=\go ^2/a^2}
  e_{S,j} (x,{\bf k}) (e_{S,j} (y,{\bf k}))^\star
 d\gs \right)
 ~,\end{array} \]
with $\gg_d (\go)$ defined by Equation (\ref{equ:gamma}).

This formula expresses the fact that the correlation of scattered 
plane waves randomized with the appropriate weights is proportional to the
Green's tensor.  

\end{itemize}

\section{Appendix: A short review about
 pseudo-differential operators}\label{sec:psido}

We will define pseudo-differential operator ($\Psi$DO's) on $\R ^d$.
$\Psi$DO's on manifold are defined locally by the same formulae.
More details can be foud in \cite{YCdV1,Di-Sj,Du,Tr}.
\begin{defi}
\begin{itemize}
\item 
The space $\Sigma _k $ is the space of smooth functions
$p: T^\star \R^d \ra \C $ which satisfies
\[ \forall \ga ,\gb ,~ | D^\ga_xD^\gb _\xi p(x,\xi)|
\leq C_{\ga,\gb}(1+|\xi|)^{k-|\gb|}~.\]
\item 
A symbol of order $m$ and degree $l$ is a family of functions
\[ p_\ge : T^\star \R^d \ra \C \]
which admits an asymptotic expansion
\[ p_\ge \sim \sum _{j=0}^\infty \ge ^{j+m} p_{j}(x,\xi) \]
with $p_j \in \Sigma ^{l-j}$. We will denote this space by
$S_{m,l}$.
\end{itemize}
\end{defi}

\begin{defi}
An  $\ge$-pseudo-differential operator $P$ (a $\Psi DO$)
of order $m$ and degree $l$
on $\R^d$ is given locally by the kernel
\[ [P](z,z')=(2\pi \ge  )^{-d}\int _{\R^d}e^{i\langle  z-z'|\zeta
 \rangle /\ge   }p_\ge \left(\frac{z+z'}{2}, \zeta \right)
|d\zeta |\]
where  $p_\ge (z, \zeta) $, the so-called {\it (total) symbol} of $P$,
is in $S_{m,l}$.
\end{defi}

We will denote 
$ P={\rm Op}_\ge  (p_\ge )$.

The kernel of $P$ is then given by:
\[ [P](z,z')=\ge ^{-d}\tilde{p}\left( \frac{z+z'}{2} ,\frac{z'-z}{\ge}\right)
 \]
with $\tilde{p}$ the partial Fourier transform of $p_\ge (z,\zeta)$ w.r. to 
$\zeta $.
Very often, one is only able to compute the symbol  $p_0$
which is called the {\it principal symbol} of ${P}$.

The most basic fact about $\Psi DO$'s is the fact they can be composed:
if $P={\rm Op}_\ge (p)$ and $Q={\rm Op}_\ge (q)$, we have
$PQ={\rm Op}_\ge \left( pq +O(\ge) \right)$.

\bibliographystyle{plain}

\begin{thebibliography}{99}

\bibitem{Bo-Ro}
A. Bouzouina \& D. Robert.
Uniform semi-classical estimates for the propagation
of quantum observables.
\newblock{\em Duke Math. Jour.,}
111(2):223--252 (2002).





\bibitem{DL}
  A. Derode, E. Larose, M. Tanter, J. de Rosny, A. Tourin,
M. Campillo \&  M. Fink,
\newblock Recovering the Green's   function  from field-field
 correlations in an open
 scattering medium.
 \newblock{\em  J. Acoust. Soc. Am.,}
113:2973-2976 (2004).

\bibitem{DLC}
  A. Derode, E. Larose, 
M. Campillo \&  M. Fink,
\newblock How to estimate the Green's   function of a heterogeneous
medium between to passive sensors? Application to acoustic waves.
\newblock{\em Applied Physics Lett.,}
83(15):3054-3056 (2003).

\bibitem{YCdV1}
Y. Colin de Verdi\`ere,
{M\'ethodes semi-classiques et th\'eorie spectrale.}
\newblock{\em Lecture Notes in Preparation}
\newblock{\it (
\textcolor{blue}{http://www-fourier.ujf-grenoble.fr/${\rm \tilde{~} }$ycolver/ }~)}

\bibitem{YCdV2}
Y. Colin de Verdi\`ere,
Mathematical models for passive imaging\\
II: Effective Hamiltonians associated to surface waves,
\newblock{\em In Preparation}.


\bibitem{YCdV3}
Y. Colin de Verdi\`ere,
Mathematical models for passive imaging \\
III: Noise located on the boundary.
\newblock{\em In Preparation}.

\bibitem{Di-Sj} M. Dimassi \& J. Sj\"ostrand.
Spectral Asymtotics in The Semi-Classical Limit.
\newblock{\em Cambridge Univ. Press}, 1999.

\bibitem{Du}
J. Duistermaat,
\newblock Fourier Integral Operators.
\newblock{\em Birkh{\"a}user}, Boston, 1996.

 \bibitem{GV}
I.M. Gelfand \& N.Y. Vilenkin.
\newblock{Les distributions IV : applications de l'analyse harmonique.}
\newblock{\em Dunod (Paris),}
 1967.



\bibitem{Hor}
L. H\"ormander,
\newblock The spectral function of an elliptic operator.
\newblock{\em Acta Mathematica,}
121, 193--218 (1968).







\bibitem{LW}  O.I. Lobkis \& R.L. Weaver,
\newblock
On the emergence of Green's   function in the correlations of a diffuse field.
\newblock{\em  J. Acoust. Soc. Am.,}
{ 110:3011-3017 (2001)}.


\bibitem{RS3}
M. Reed \& B. Simon,
\newblock{\em Methods of modern Math. Phys. III.}



\bibitem{RSK} Ph. Roux, K. Sabra \& W. Kupperman,
\newblock
Ambient noise cross correlation in free space: Theoretical approach.
\newblock{\em  J. Acoust. Soc. Am.,}
{ 117:79-84 (2005)}.

\bibitem{SS} F. Sanchez-Sesma, J. P\'erez-Ruiz, M. Campillo \&
F. Luz\'on,
\newblock{The elastodynalic 2D Green function retrieval from cross-correlation:
the canonical inclusion problem.}
\newblock{\em Preprint,}
(2006).


\bibitem{SC}
N. Shapiro \& M. Campillo,
\newblock{Emergence of broadband Rayleigh waves from
 correlations of the ambient
 seismic noise.}
\newblock{\em Geophys. Res. Lett.,}
31:L07614 (2004).

\bibitem{SCSR}
N. Shapiro, M. Campillo, L. Stehly \& M. Ritzwoller,
\newblock High Resolution Surface Wave Tomography From Ambient Seismic Noise.
\newblock{\em Science,}
307:1615  (2005).





\bibitem{SRTDHK}
K. Sabra, P. Roux, A. Thode, G. D'Spain, W. Hogliss \&
W. Kuperman,
\newblock{Using Ocean Ambient Noise for Array Self-Localization
and Self-Synchronization.}
\newblock{\em IEEE J. of Oceanic engineering,}
30:338-346 (2005).

\bibitem{Schwartz}
L. Schwartz.
\newblock{Radon measures on arbitrary topological
 spaces and cylindrical measures.}
\newblock{\em Oxford University Press,}
1973.

\bibitem{Tr}
{F. Tr{\`e}ves,}
\newblock {Introduction to pseudodifferential and {F}ourier integral
              operators.}
\newblock{\em Plenum Press,}
{New York}, {1980}.

\bibitem{WL1}
R.L. Weaver \& O.I. Lobkis,
\newblock
On the emergence of Green's function in the correlations
 of a diffuse field: pulse 
echo using thermal phonons.
\newblock{\em Ultrasonics,}
40, 435-439 (2002).

\bibitem{WL2}
R.L. Weaver \& O.I. Lobkis,
\newblock  Ultrasonics without a source:
 thermal fluctuation correlations at MHz 
frequencies.
\newblock{\em Phys. Rev. Lett.,}
87(13):134301-134304 (2001).

\bibitem{WL3}
R.L. Weaver \& O.I. Lobkis,
\newblock Diffuse fields in open systems and the emergence of the
Green's functions (L).
\newblock{\em J. Acoust. Soc. Am.,}
116(5):1-4 (2004).

\bibitem{We}
R.L. Weaver,
\newblock Information from seismic noise.
\newblock{\em Science,}
307:1568-1569 (2005).


\bibitem{Wi}
N. Wiener,
\newblock Extrapolation, Interpolation and Smoothing of
 Stationary Time Series.
\newblock{\em MIT and  John Wiley,}
1950.



\end{thebibliography}

\end{document}